# *Inertial Mass, Its Mechanics – What It Is; How It Operates*

*Roger Ellman*

Abstract

The subject is treated in two parts, inertial mass and gravitational mass. The present paper is the first part: *Inertial Mass, Its Mechanics – What It Is; How It Operates*.

The behavior of inertial mass is well known, described by Newton's Laws, the Lorentz Contractions, and Einstein's mass – energy equivalence. But just what mass is, how those behaviors come about, what in material reality produces the effects of inertial mass, is little understood.

The only extant hypothesis is the "Higgs Field" and its related particle, the Higgs Boson. Neither has been detected in spite of significant efforts. Further, their hypothesis is not a description of the mechanics of mass but an abstraction away from the problem, substituting another field to explain that not well understood.

From a start of only the limitation on the speed of light, the necessity of conservation, and the impossibility of an infinity in material reality, the present paper presents a comprehensive analysis of the phenomenon inertial mass:

- how rest mass appears in particles,

- how the Newtonian behavior arises from that, and

- how the Lorentz Contractions operate in/on it,

or, in other words, the mechanics of inertial mass.

Roger Ellman, The-Origin Foundation, Inc.
        http://www.The-Origin.org
        320 Gemma Circle, Santa Rosa, CA 95404, USA
        RogerEllman@The-Origin.org

# Inertial Mass, Its Mechanics – What It Is; How It Operates

## Roger Ellman

### 1. DEVELOPMENT OF THE UNIFIED FIELD

The following "thought experiments" develop the concept.

Electric Field

- Nothing can travel faster than the speed of light, $c$. Given two static electric charges separated and with the usual Coulomb force between them, if one of the charges is moved the change can produce no effect on the other charge until a time equal to the distance between them divided by $c$ has elapsed.

- For that time delay to happen there must be something flowing from the one charge to the other at speed $c$ and the charge must be the source of that flow.

    The Coulomb effect is radially outward from the charge, therefore every charge must be propagating such a flow radially outward in all directions from itself, which flow must be the "electric field".

Motion of Charge and "At Rest"

- Comparing two such charges, one moving at constant velocity relative to the other, at least one of the charges is moving with some velocity, $v$.

- The flow (of "field") outward from that charge must always travel at $c$. Forward it would go at $[c + v]$ if propagated at $c$ from the source charge already moving that way at $v$. Therefore, it must be sent forward from the charge at $[c - v]$ so that it will travel at $c$ when the $v$ of its source charge is added.

*(1)*   $\text{Flow}_{fwd} = [c - v]$

- Analogously, rearward it would go at speed $[c - v]$ if propagated at $c$ from the source charge already moving the opposite way at $v$. Therefore, it must be emitted rearward from the charge at $[c + v]$ so that it will travel at speed $c$ when the $v$ of the source charge in the opposite direction is subtracted.

*(2)*   $\text{Flow}_{rwd} = [c + v]$

- But, that rearward – forward differential means that the direction and speed of motion can be determined by looking at the propagation pattern of the flow as propagated by the charge.

    And, if the pattern were the same in all directions then the charge would be truly "at rest", which means that there is an absolute "at rest" frame of reference.

Unification of Fields

- Except for the kind of field, all of the preceding applies in the same way and with the same conclusions for magnetic field and gravitational field as for electric field.

- Therefore, either a particle that exhibits all three such fields, as for example a proton or an electron, is a source of three separate and distinct such flows, one for each field, or there is only a single flow which produces all three effects: electric, magnetic, and gravitational.

    The only reasonable conclusion is that electric, magnetic, and gravitational field are different effects of the same sole flow from the source particles.

Sources & Their Decay

- The flow is not inconsequential. Rather, it accounts for the forces, actions and energies of our universe.

- For a particle to emit such a flow the particle must be a source of whatever it is that is emitted outward. The particle must have a supply of it.

- The process of emitting the flow from a particle must deplete the supply resource for the particle's emitting further flow, must use up part of its supply, else we would have something-from-nothing and a violation of conservation.

    It must be concluded that an original supply of that which is flowing came into existence at the beginning of the universe and has since been gradually being depleted at each particle by its on-going outward flow.

That Which is Flowing

- The flow is a property of contemporary particles. Those particles are evolved successors to the original oscillations with which the universe began. Then, that which is flowing is the same original primal "medium", the substance of the original oscillations at the beginning of the universe.

    Since it is flowing outward from the myriad particles of the universe simultaneously and that flow is interacting with myriad others of those particles without untoward interference, the "medium" must be extremely intangible for all of that to take place, any one particle's flow flowing largely freely through that of other particles, as intangible as -- well -- "field".

The Beginning

- Before the universe began there was no universe. Immediately afterward there was the initial supply of medium to be propagated by particles. How can one get from the former to the latter while: (1) not involving an infinite rate of change, and (2) maintaining conservation ?

    The only form that can accommodate the change from nothing to something in a smooth transition without an infinite rate of change is the oscillatory form of equation *(3a)*, below.

*(3a)*   $U_0 \cdot [1 - \text{Cos}(2\pi \cdot f \cdot t)]$

    The only way that such an oscillation can have come into existence without violating conservation is for there simultaneously to have come into existence a second oscillation, the negative of equation *(3a)* as in equation *(3b)*.

*(3b)*   $-U_0 \cdot [1 - \text{Cos}(2\pi \cdot f \cdot t)]$

    That is, the two simultaneous oscillations must have been such as to yield a net of nothing, the prior starting point, when taken together.

The Oscillatory Medium Flow ≡ Electric charge and field

- The initial medium supply of each particle, each being a direct "descendant" of the original oscillation at the universe's beginning, must be oscillatory in form per equations *(3)*. Therefore the radially outward flow from each particle is likewise an oscillatory medium flow of the form of equations *(3)*.

    The flow is radially outward from the particle, therefore, the oscillation of the medium supply of each particle is a spherical oscillation. The particle can also be termed a *center-of-oscillation*, which term will also be used here.

- The amplitude, $U_0$, of the *[1 - Cosine]* form oscillation is the amplitude of the flow emitted from the source particle, which flow corresponds to the electric field. Thus the oscillation amplitude must be the charge magnitude of the source particle -- the fundamental electric charge, $q$, in the case of the fundamental particles, the electron and the proton.

> Then, the conservation-maintaining distinction of amplitude $+U_0$ versus amplitude $-U_0$ must be the positive / negative charge distinction.
>
> The frequency, $f$, of the $[1 - Cosine]$ form oscillation must then correspond to the energy and mass of the source particle, that is the energy of the oscillation is $E = h \cdot f$ and the mass is $m = E/c^2 = h \cdot f/c^2$.

[- While it does not pertain to the universe's beginning, because the outward medium flow from each particle must deplete the source particle's remaining supply of medium for further propagation, the amplitude magnitude, $U_0$, must exponentially decay. That is, it must be of the form of equation $(4)$, below.]

$$(4) \quad |U(t)| = U_0 \cdot \varepsilon^{-t/\tau}$$

## Medium Emission and Medium Flow

- When a charge is at rest, medium is emitted by it and flows outward in the same manner in all directions. But, when the charge is in motion at constant velocity, $v$, the flow forward is emitted at speed $[c - v]$ and rearward at $[c + v]$ per above.

- There can be only one frequency, $f$, in the $[1 - Cosine(2\pi \cdot f \cdot t)]$ form oscillation of the emitted flow regardless of whether it is directed forward, rearward or sideward. Therefore, to obtain the slower speed, $[c - v]$, emitted forward the wavelength forward, $\lambda_{fwd}$, must be shorter so that the speed at which the flow is <u>emitted</u>, $= f \cdot \lambda_{fwd}$, will be slower.

- The case is analogous rearward where $\lambda_{rwd}$ is longer in order for the speed, $[c + v]$, to be greater.

> In all directions from the moving charge, including any that are partially sideward plus partially forward or rearward, the speed of emission and the wavelength emitted will be the resultant of the sideward plus forward or rearward components of a ray in that direction.

- The absolute rate of flow outward of the emitted medium must be at speed $c$. Forward that comes about because the forward speed of the charge, $v$, adds to the forward speed at which the medium is emitted, $[c - v]$, resulting in the medium flowing at the speed of the sum, $speed = v + [c - v] = c$.

- That speed increase raises the $[1 - Cosine(2\pi \cdot f \cdot t)]$ form oscillation frequency (per the Doppler Effect). Thus forward medium <u>flow</u> speed is $c = f_{fwd} \cdot \lambda_{fwd}$.

- Analogously rearward the speed of medium <u>flow</u> is at $c = f_{rwd} \cdot \lambda_{rwd}$.

> In all directions from the moving charge, including any that are partially sideward plus partially forward or rearward, the speed of flow will be $c$ and the frequency and wavelength of the flow will be the resultant of the sideward plus forward or rearward components of a ray in that direction.

## Magnetic Field

- A charge at rest exhibits the electrostatic effect but not the magnetic effect. That charge has a spherically uniform pattern of $[1 - Cosine(2\pi \cdot f \cdot t)]$ form oscillatory medium emission and flow outward.

- A charge in motion exhibits the magnetic effect in addition to the electrostatic effect. That charge has a pattern of emission and outward flow of medium that is cylindrically symmetrical about the direction of motion but that varies in wavelength and frequency from $f_{fwd} \cdot \lambda_{fwd}$ forward to $f_{rwd} \cdot \lambda_{rwd}$ rearward.

The electrostatic [Coulomb's Law] effect is due to charge location. The magnetic [Ampere's Law] effect is due to charge motion. Clearly, then, the electrostatic effect is due to the spherically uniform medium flow from the charge and the magnetic effect is due to the change in shape of that medium flow pattern caused by the charge's motion.

Electro-magnetic Field

- There is a continuous emission of medium in $[1 - Cosine(2\pi \cdot f \cdot t)]$ oscillatory form from each charge, which medium flows outward, away, forever. Constant velocity motion of a charge produces a change in the frequency and wavelength of that medium flow.

- Changes in the velocity of the charge cause corresponding further changes in the medium's oscillatory form as successive increments of medium are emitted and flow outward from the charge. Earlier increments so changed propagate on outward away from the charge, forever, at $c$. The stream of outward flowing medium carries a history of the motions of the source charge.

> Propagating electromagnetic field is the carrying of both of those field aspects as an imprint on the otherwise uniform medium flow from the charged particle, an imprint analogous to the modulation of a carrier wave in radio communications.

- Electro-magnetic field is caused by acceleration / deceleration of charge, that is by changes in the charge velocity. Therefore:

> The changing electric and magnetic fields of electro-magnetic field actually <u>are</u> form changes imprinted onto the outgoing medium flow and carried passively with it [analogous to modulation of a carrier wave in radio communications].

> Because all medium flow is spherically outward in all directions from its source charge, changes in it, caused by changes in the source charge velocity, propagate outward <u>in all directions</u>. Those medium flow changes <u>are the</u> changing electric and changing magnetic fields of <u>electro-magnetic field.</u>

- It is not the speed of light which is the fundamental constant, $c$, light being a mere modulatory imprint on medium flow. It is the speed of medium flow which is the fundamental constant, $c$.

Gravitational Field

- As pointed out earlier above, the frequency, $f$, of the $[1 - Cosine]$ form oscillation corresponds to the mass of the source particle. Therefore the frequency aspect of the radially outward medium flow is the "gravitational field". See[3] for more on gravitation.

## 2. *MASS AND COULOMB'S LAW*

Just as the $[1 - Cosine]$ type oscillatory wave of propagated medium is the field, so the source of that propagation, the source itself from which the propagation is emitted radially outward in all directions, [hereafter referred to as *center-of-oscillation*] must embody the charge and the mass, that is, the matter of the "particle" involved. Those are the only physical realities underlying our entire universe: the center-of-oscillation and the propagated wave of medium. Those alone are, cause, account for all of matter, mass, field, force, charge, energy, radiation, everything.

The effect called "force" is, then, the result of the waves propagated by a center- or centers-of-oscillation arriving at and interacting with an encountered center, the center upon which the force is exerted. (In the following discussion, centers will be referred to as the "source" center and the "encountered" center. Of course every center is continuously in both roles. The distinction here is only in order to clarify the discussion. For this purpose, subscript "e" signifies encountered center and subscript "s" signifies source center.)

The effect of an individual cycle of wave encountering a center is the delivery to the center of an *impulse* [an impulse is a *force × time*], which is an amount of *momentum* *change*. The wave cycle, then, as it is propagated by its source center, carries potential impulse, "potential" because it is not realized in an effect until an encounter with another center occurs. The amount of potential impulse in the wave is, of course, proportional to the amplitude of the wave. It is that amount, that amplitude, which decreases as the square of the distance from the source center because it becomes spread over a greater area. The overall stream of successive wave cycles carries the potential impulse of one wave times the repetition rate, the frequency, of the waves.

The potential status of the wave impulse is exactly the same status as that of electric field (which it, in fact, is) where electric field is potential force and not realized as actual force until it interacts with an electric charge (a center-of-oscillation). That unrealized, potential, effect of the incoming waves from a source center, the waves' magnitude multiplied by their repetition rate or frequency, then becomes realized as actual effect by the interaction with an encountered center.

Newton's Law,

*(5)* Force = Mass × Acceleration

can be restated as

*(6)* Acceleration Resulting = Force Applied × $\frac{1}{Mass}$

and in that form is a more natural statement since force is the cause and acceleration the effect. This translates in terms of waves and centers into

*(7)* $\begin{bmatrix} \text{Acceleration} \\ \text{Resulting} \end{bmatrix} = \begin{bmatrix} \text{Wave} \\ \text{Potential} \\ \text{Impulse} \end{bmatrix} \times \begin{bmatrix} \text{Responsiveness} \\ \text{of the Center} \end{bmatrix}$

or, more succinctly,

Acceleration = Wave × Responsiveness.

While the "acceleration" quantity of Newton's Law is retained in the revised formulation of equation *(7)*, the other quantities are not directly analogous. That is, the expression of the acceleration in the product of two components as in equation *(7)* involves a different pair of components than those of equations *(5)* and *(6)*. Force is not the same as "wave potential impulse" nor is "responsiveness of the center" identical to inverse mass. This distinction will be clarified shortly.

[In actual events the resulting acceleration may also result in changes in the incoming wave (by motion of the center changing the separation distance with the consequent inverse square law effect on wave magnitude, for example) and it may result in changes in the encountered center's responsiveness (to be seen shortly and the equivalent of relativistic changes in mass with velocity).]

In this formulation Responsiveness appears in the role of $^1/_{Mass}$ so that it would appear that

*(8)* Responsiveness $\propto \frac{1}{Mass}$  or  Mass $\propto \frac{1}{Responsiveness}$

however, the connection is more subtle as will be seen shortly.

The responsiveness would seem to depend upon several factors. The first factor is encountered center cross-section, the effective "target" area that the encountered center has for intercepting incoming waves. Of the total wave traveling outward from the source center, the only part that interacts with another center is the part that encounters the center, that is intercepted by the encountered center. A center intercepting a "large" portion of wave would receive a greater impetus to change in motion from the arriving wave than would a center intercepting a "small" portion of the wave. All other things being the same, the center intercepting the larger portion of incoming wave would receive the greater impulse and consequently would experience the greater momentum change. Thus inverse inertial mass, or center responsiveness, must depend on the encountered center's cross-section for interception of medium waves as one factor.

The discussion here assumes that the arriving wave comes from a sufficiently distant center that the wave is effectively a plane wave, that is, that the part of the wave intercepted by the encountered center is a flat wave front of which every part travels parallel to the center part. [The non-plane wave case is treated elsewhere[2] and turns out to be of negligible effect even with the relatively non-distant separation of the orbital electrons of atoms from the nucleus except a slight effect, the "Lamb Shift", detectable at the closer orbits.]

A center of smaller cross-section is of greater mass (lesser responsiveness), all other effects being equal, and requires greater arriving wave amplitude to experience a specified change in motion than does a center of larger cross-section. Cross-section is a matter of size, that is it is proportional to the area of interception of the incoming wave front. The encountered center being a spherical oscillation the cross-section is the area of a circle perpendicular to the direction of travel of the wave front as it encounters the center. That area should depend on the wavelength of the encountered center's oscillation; more precisely it should be proportional to the square of that center's wavelength since the area of the circle will be $\pi$ (pi) times the square of the radius.

One cannot make a statement as to the definite "size" of the center in space since it is continuously varying; however it is convenient to think of its "size" as a sphere of radius equal to the wavelength. Since for the moment the cross-section is being taken as proportional to the square of the wavelength, not equal to it, there is no problem that the "size" is not determinate.]

This yields the first factor in mass, or responsiveness,

*(9)*    Cross-section  $\propto$  $\pi \cdot \lambda_e^2$  =  $K_{cs} \cdot \lambda_e^2$

*(10)*   Responsiveness $\propto$ [Factor 1]·[Factor 2]·[Factor 3]

$\propto$ [ Cross-section ]·[  "  ]·[  "  ]

= [ $K_{cs} \cdot \lambda_e^2$ ]·[  "  ]·[  "  ]

where:  $K_{cs}$ = a constant for the proportionality
        $\lambda_e$  = the wavelength of the encountered
                 center oscillation

In order to account for the action of waves on an encountered center relating only to that portion of the total wave front intercepted by the encountered center, the incoming wave must be expressed in terms of "Incoming Wave Potential Impulse per Unit Area". That is, the intercepted wave potential impulse, the "Wave" of equation *(7)* is

*(11)*
$$\text{Wave} = \frac{\text{Total Propagated Wave Potential Impulse of Source Center}}{\text{Total Spherical Area of Source Wave at Distance Encountered Center is from Source}}$$

= Wave Potential Impulse per Unit Area

so that upon being multiplied by the cross-sectional area at the encountered center the units of area are cancelled and the resulting quantity is wave impulse (as measured at and as intercepted by the encountered center). The division by the area of a sphere is the essence of the inverse square law, of course.

The second factor in responsiveness is the effective amplitude of the encountered center's oscillation during the interaction. The medium wave being a *[1 - Cos]* form as in Figure 1, below, a range of possible interactions can occur because of various different potential source and encountered center frequencies and phases. The extremes and mean of the range of possible encounters are as follows.

(1) Frequency$_{wave}$ << Frequency$_{center}$

If the frequency of the arriving wave is much less than that of the encountered center then the encountered center goes through all of its amplitude values, as above, many times while one wave arrives. Clearly in this case the effective amplitude during the encounter with that one wave is the average amplitude of the encountered center's oscillation.

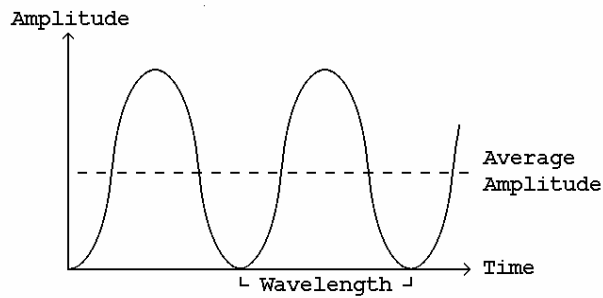

*Figure 1*
*Oscillation of a Center*

(2) Frequency$_{wave}$ >> Frequency$_{center}$

If the frequency of the incoming wave is much greater than the frequency of the encountered center then one arriving wave encounters the center while the center is essentially at one value of its oscillation (wherever on the curve of Figure 1 that it happened to be). However, successive incoming waves will encounter the center at various other points on the curve and the average of a large number of such encounters will again be the average amplitude of the encountered center's oscillation.

Since the frequencies are very large this averaging is valid unless one frequency is very close to being a multiple of the other and the phases do not vary. However, the frequencies are constantly changing due to the effects that produce relativistic mass changes and the phase is constantly changing because of relative motion of each center with regard to the other.

Furthermore, in real matter, not our idealized model of one source and one encountered center, every center is constantly "bombarded" by various waves from a variety of directions at a variety of frequencies due to the immense number of centers making up ordinary matter. These can be analyzed individually in accordance with the Fourier principal; however, the total effect is such that it would be an extremely rare event for a center to maintain both phase and frequency-multiple relationships with a wave for even one cycle.

(3) Frequency$_{wave}$ = Frequency$_{center}$

Finally, if the frequencies are the same then the interaction takes place over exactly one cycle and the effective amplitude is, again, the average.

Thus the relative frequency and the relative phase of the wave and the center that it encounters have no effect on the large scale result from the interaction. The `Factor 2` is not a variable quantity but merely the average amplitude of the encountered center, which is now designated $U_e$.

However, the absolute frequency of the encountered center is very applicable and yields the third factor in the formula for responsiveness. Just as the incoming wave repetition rate affects the amount of force that the wave can deliver to the encountered center, as already presented, so the encountered center repetition rate affects that center's responsiveness to the wave. While the wave is encountering the center, each cycle of the encountered center's oscillation is affected by the wave. The encountered center is acted upon by the incoming wave each time the center oscillates out from the zero point of its oscillation to its peak, which occurs at the frequency of the encountered center. (This is most easily visualized if the frequency of the encountered center is much larger than that of the wave, but it applies in any case.)

The *Factor 3*, then, is the encountered center frequency and equation *(10)* becomes

*(12)* Responsiveness = [ Cross-section ]·[Amplitude]·[Frequency]

$$= [\ K_{cs} \cdot \lambda_e^2\ ] \cdot [\ U_e\ ] \cdot [\ f_e\ ]$$

where: $K_{cs}$ = a constant for the proportionality
$\lambda_e$ = the wavelength of the encountered center oscillation
$U_e$ = its amplitude, and
$f_e$ = its frequency

Since the wave propagates at the speed of light, c, the frequency and the wavelength are related as $f = c/\lambda$, which substituted into equation *(12)* yields

*(13)* Responsiveness = $K_{cs} \cdot \lambda_e^2 \cdot U_e \cdot f_e$ = $K_{cs} \cdot \lambda_e^2 \cdot U_e \cdot [c/\lambda_e]$

$$= K_{cs} \cdot \lambda_e \cdot U_e \cdot c$$

Further, recognizing that $K_{cs}$ is a constant for the proportionality, $c$ is a constant, and $U_c$ is a constant quantity [yet to be specified], then mass and responsiveness are related proportionally as in equations *(14a)* and *(14b)*, below.

*(14)*     (a)                                                                          (b)
Responsiveness ∝ $\lambda_e$ ∝ $1/f_c$           Mass ∝ $1/\lambda_e$ ∝ $f_c$

Thus the responsiveness of a center to change in its motion due to a wave arriving and encountering it is directly proportional to the center's wavelength and inversely proportional to its frequency. The mass is directly proportional to the center frequency and inversely proportional to the wavelength.

This result is also given by traditional 20th Century physics (if a new, but reasonable assumption is introduced) as follows.

(1) From traditional 20th Century physics the energy equivalent of mass is

*(15)*   $E = m \cdot c^2$

(2) From traditional 20th Century physics the energy equivalent of an oscillation is

*(16)*   $E = h \cdot f$

(3) Applying these to a center-of-oscillation that represents a particle of mass *m*, and is oscillating at frequency *f*, there can be only one total energy of the particle / center, therefore,

*(17)*   $m \cdot c^2 = h \cdot f$

$m \cdot c \cdot [\lambda \cdot f] = h \cdot f$

$$m = \frac{h/c}{\lambda}$$

But, $h/c$ is the ratio of two universal constants and, therefore, a constant itself. Thus mass is inversely proportional to center wavelength and directly proportional to center frequency -- the same result as obtained by analyzing the wave interaction with the encountered center.

The "reasonable new assumption" is that Planck's Constant, $h$, may be used with a center-of-oscillation's frequency to get the energy equivalent as in step (2), above, just as for a photon the energy is its frequency (electromagnetic wave frequency) multiplied by $h$, which has been long established. That the result produces the same relationship between mass and frequency would justify the assumption.

Taking the values listed below for the quantities involved (they are known to much greater accuracy but as below is sufficient for the moment),

```
h = 6.63×10⁻²⁸ erg-sec              c = 3.00×10¹⁰ cm/sec
m_e- = 9.17×10⁻²⁸ gm [electron]     m_p+ = 1.68×10⁻²⁴ gm [proton]
```

the frequency and wavelength of the electron and the proton are as follows.

```
f_e- = 1.24×10²⁰ Hz                 f_p+ = 2.27×10²³ Hz
λ_e- = 2.41×10⁻¹⁰ cm                λ_p+ = 1.32×10⁻¹³ cm.
```

Experiments involving scattering of charged particles by atomic nuclei have yielded an empirical formula for the approximate value of the radius of an atomic nucleus to be

*(18)*    `R = (1.2×10⁻¹³) × (Mass Number, A) cm`

The proton is the nucleus of the Hydrogen atom with mass number `A=1`. For that value equation *18* gives a radius of `R=1.2×10⁻¹³cm`. That result is quite close to the value of $\lambda_{p+}$ just obtained above.

It is a curious result, but nevertheless the case, that whereas we tend to think of the electron as "small" and the proton as "large" (because the electron mass is much smaller than the proton mass), actually the electron appears "large" relative to the proton. The reason is that it requires larger cross-section, greater responsiveness, to have smaller mass. Thus the electron with a rest mass about $1/1836$ of the proton rest mass has a wavelength *1836* times that of the proton and an implied "radius" *1836* times that of the proton. Contrary to the traditional schematic diagram or animated illustration showing a hydrogen atom with a large proton nucleus and a small orbital electron, and contrary to the large sun and small planets analogy, the actual case would appear to be of a large electron orbiting a small nucleus.

Whether one could ever define or identify a "radius" for a center-of-oscillation is open to question. The concept of radius assists in thinking about centers-of-oscillation but is not otherwise applicable to the physics discussion. Cross-section, which is very applicable to the physics, need not at all be the same magnitude as the area of the circle of radius equal to the center's "radius". But, nevertheless, the relative cross-sections of different centers must be proportional to the square of their relative wavelengths.

Likewise, the question of the actual interaction of wave with center requires care. It is tempting to think in macroscopic physical terms of impulse, force, "bumping" and so on. The principle of equivalence would say, "These actions and events appear to be like the macroscopic Newtonian actions and events and they in fact are the actual "microscopic" actions and events that produce the macroscopic Newtonian ones, therefore they must be the same." The seeming defect in equivalence is in responsiveness versus mass and in waves and centers versus "hard" physical objects. On the other hand, that is the entire point of the principle of equivalence, and it is correctly applied here (and is more comfortable) if the order of statements is reversed:

- not "Wave - center interactions are the same as Newtonian force, impulse, etc. type actions and events", but rather

- "Newtonian force, impulse, etc. type actions and events are really wave - center interactions,"

which is the case.

Consequently, it is clearer and more straight forward to use the Newtonian terms of force, impulse, etc. in describing wave - center behavior.

### 3. PRECISE FORMULATION AND COULOMB'S LAW

In the traditional formulation of Newton's Law as inverted (equation *(6)* repeated here)

*(6)* $\text{Acceleration Resulting} = \text{Force Applied} \times \dfrac{1}{\text{Mass}}$

and for the case that is now being considered, that in which the force results from the electrostatic interaction between two charges in accordance with Coulomb's Law

*(19)* $\text{Force} \propto \dfrac{\text{Charge} \times \text{Charge}}{[\text{Separation Distance}]^2}$

the charge of both of the interacting centers enters into the Newton's Law relationship in the Force part, the Mass part of the relationship being like an inert characteristic of the substance.

In the new formulation (equation *(7)*, repeated here)

*(7)* $\begin{bmatrix}\text{Acceleration}\\ \text{Resulting}\end{bmatrix} = \begin{bmatrix}\text{Wave}\\ \text{Potential}\\ \text{Impulse}\end{bmatrix} \times \begin{bmatrix}\text{Responsiveness}\\ \text{of the Center}\end{bmatrix}$

or, more succinctly,

$\text{Acceleration} = \text{Wave} \times \text{Responsiveness}.$

the quantity that is called oscillation amplitude, $U$, the role of which corresponds to that of traditional charge, $Q$, enters into the formulation somewhat differently. The source center's amplitude is a factor in the Wave and the encountered center's amplitude is a factor in the Responsiveness.

While the traditional formulation gives correct answers, it is incorrect as a reflection of the process or events occurring. While the electrostatic effect interaction magnitude is mutual and both charge magnitudes enter into the effect, they do not do so in the same way. Each is a propagator of waves and is an encountered center. Those two actions are not identical. The process is better described by the formulation that embodies the two aspects, Wave and Responsiveness, in which formulation the source center amplitude / charge enters into the Wave factor and the encountered center amplitude / charge enters into the Responsiveness factor.

Thus as here formulated, responsiveness is not simply the inverse of mass since it includes the encountered center's amplitude / charge. Figure 2, below, presents a comparison of the two methods of viewing the action and makes clear that the difference is (in 20th Century physics terms) whether the resulting acceleration is expressed in terms of applied force or applied field.

| 20th Century Physics | Revised Point of View |
|---|---|
| $\text{Acceleration} = \text{Force} \times \dfrac{1}{\text{Mass}}$ | $\text{Acceleration} = \text{Wave} \times \text{Responsiveness}$ |
| $= \dfrac{Q \cdot Q}{d^2} \times \dfrac{1}{\text{Mass}}$ | : |
| $= \dfrac{Q_s}{d^2} \times \dfrac{Q_e}{\text{Mass}}$ | substituting for responsiveness with equation *(13)* |
| $= \begin{bmatrix}\text{Electric}\\ \text{Field at}\\ \text{Radius d}\end{bmatrix} \times \dfrac{Q_e}{\text{Mass}}$ | $= \text{Wave} \times \begin{bmatrix}K_{cs} \cdot \lambda_e \cdot U_e \cdot c\end{bmatrix}$ |

*Figure 2*

Field and Wave, not Force and Wave, correspond. Each is the unrealized potential that becomes action via interaction with an encountered *charge / center*. Therefore the *charge / mass* of the left half of Figure 2 is the same as the *responsiveness* of the right half of the figure as follows (with letting $m_e$ symbolize the encountered mass).

*(20)*
$$\frac{Q_e}{m_e} = K_{cs} \cdot \lambda_e \cdot U_e \cdot c$$

And since, from equation *(17)*, rearranged

*(17)*
$$m_e = \frac{h}{\lambda_e \cdot c}$$

then

*(21)*
$$Q_e = \frac{h}{\lambda_e \cdot c} \cdot [K_{cs} \cdot \lambda_e \cdot U_e \cdot c]$$

$$= h \cdot K_{cs} \cdot U_e$$

which relates the charge of the encountered center to that center's amplitude, and is a simple direct proportionality because $h$ and $K_{cs}$ are constants.

If time could be stopped so that the waves were frozen in whatever position that they had in space, then the spherical waves as propagated by a center would appear as a series of successive nested shells, each of a successively greater radius, the radius being

*(22)* $\quad R_w = n \cdot \lambda_w$

```
            where:   n  = 1, 2, 3 ... for the successive shells
                     λ_w = the wavelength of the waves
```

and the thickness of each shell is the wavelength, $\lambda_w$. One such shell is depicted two-dimensionally in Figure 3, below.

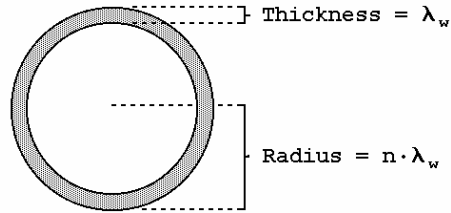

*Figure 3*

If the thickness is much less than the radius then numerical accuracy is not significantly affected by whether the radius is measured to the inner or the outer edge of the shell or to half way between, it can be taken as simply $n \cdot \lambda_w$.

As already presented, the wave propagated by the center-of-oscillation and the center's oscillation itself are of a *[1-Cosine]*. That is, the wave and center are of the forms

*(23)* $\quad$ Wave $= U_w \cdot [1 - \text{Cos}(2\pi \cdot f \cdot t)]$

$\quad\quad\quad$ Center $= U_c \cdot [1 - \text{Cos}(2\pi \cdot f \cdot t)]$

A cross-sectional view of this wave in space, that is a graph of its amplitude variation along a radius while traversing the thickness, is depicted in Figure 4, below, from which it is clear that the area under the curve of amplitude variation is equal to $U_w \cdot \lambda_w$. (For large $n$ the inverse square radial decrease in $U_w$ over one $\lambda_w$ can be neglected.)

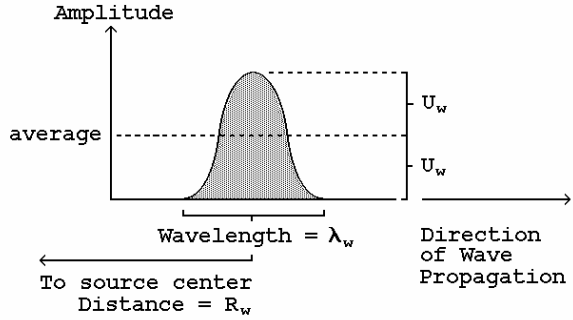

*Figure 4*

The amount of potential impulse in one complete spherical shell, in one wave cycle, is the product of the amount where a radius traverses the shell, a "shell cross-section" so to speak as obtained above from Figure 4 as $U_w \cdot \lambda_w$, multiplied by the spherical surface area of the shell. That is,

*(24)* $\begin{bmatrix} \text{A Cycle} \\ \text{of Wave} \end{bmatrix} = \begin{bmatrix} \text{Medium Density Over} \\ \text{a Shell Thickness} \end{bmatrix} \cdot \begin{bmatrix} \text{Shell Sphere} \\ \text{Surface Area} \end{bmatrix}$

$$= [U_w \cdot \lambda_w] \cdot [4\pi \cdot R_w^2]$$

This quantity does not change as the wave propagates; it is merely more dispersed or "thinned" out as the wave propagates further, as $R_w$ increases. Therefore it is the total medium density variation in any of the cycles of propagating waves (the "shells" of Figure 3) and as is propagated by any one cycle of the oscillation of the source center. Thus the wave amplitude, $U_w$, is the source center amplitude, $U_c$, spread out over the surface of the sphere of radius $R_w$, the distance of the wave from the source center.

*(25)* $$U_w = \frac{U_c}{4\pi \cdot R_w^2}$$

This is all that there is to the wave. Consequently, by substituting equation *(25)* into equation *(24)* the single wave potential impulse of the source center is obtained as

*(26)* $U_c \cdot \lambda_c$

The *[Electric Field at Radius d]* (of Figure 2, left column) or *Wave* (of Figure 2, right column) is that single wave quantity multiplied by its repetition rate, the frequency, $f$, so that the "source" charge, $Q_S$, is

*(27)* $Q_S = [U_c \cdot \lambda_c] \cdot f_c$

$\qquad\quad = U_c \cdot c$

which relates the field of the source center to that center's oscillation amplitude and, therefore, relates the charge of the source center to its amplitude.

Combining equation *(23)* for $Q_e$, the "encountered" charge, and equation *(27)* for $Q_S$, the "source charge, and recognizing that every center is always simultaneously in both source and encountered roles, then it must be concluded that every charge, $Q$, is

*(28)* $\quad Q = U \cdot c \qquad$ and $\qquad K_{cs} = c/h$

Since (per equation *(27)*) frequency and wavelength enter into the value of the charge or field of a center only as the product of the two, which product is a constant, the speed of light, $c$, the charge and field of a center are independent of the center frequency and wavelength. Thus if all centers oscillate at the same amplitude, $U_c$, then they will each be of the same charge and have the same field strength.

This total wave / field corresponds to the charge, $Q$, the constant fundamental charge of the universe, the charge of the electron and the proton, the positron and the negaproton. (The correspondence of the wave field to the charge is the same as the theorem in traditional 20th Century physics (Stokes theorem) that the integral of the electric field over a closed surface equals the enclosed charge.)

### 4. TWO SUCH CHARGES INTERACT ELECTROSTATICALLY AS FOLLOWS.

(1) The total potential force in the wave series as propagated by the source center is (from equation *(27)*)

(27) $$U_c \cdot c$$

(2) The total wave series potential force per unit area of wave front at the encountered center is the quantity of step (1) divided by the spherical surface at the encountered center where

$$R = \text{the distance between the two centers}$$

(29) $$\frac{U_c \cdot c}{4\pi \cdot R^2}$$

(3) The responsiveness of the encountered center is (from equation *(13)*)

(13) $$\text{Responsiveness} = K_{cs} \cdot \lambda_e \cdot U_e \cdot c$$

(4) The resulting acceleration is, therefore (substituting steps (2) and (3), above, into equation *(7)* per equation *(11)*)

(30) $$\text{Acceleration} = \begin{bmatrix} \text{Wave Potential} \\ \text{Impulse per Unit} \\ \text{Area at the En-} \\ \text{countered Center} \end{bmatrix} \times \begin{bmatrix} \text{Responsiveness} \\ \text{of the} \\ \text{Encountered} \\ \text{Center} \end{bmatrix}$$

$$= \frac{U_c \cdot c}{4\pi \cdot R^2} \times K_{cs} \cdot \lambda_e \cdot U_e \cdot c$$

(5) The mass of the encountered center is (from equation *(17)*)

(31) $$m = \frac{h}{c \cdot \lambda_c}$$

(6) The force is, then (substituting steps (4) and (5), above into equation *(5)*)

(32) $$\text{Force} = \text{Mass} \times \text{Acceleration}$$

$$= \left[\frac{h}{c \cdot \lambda_c}\right] \times \left[\frac{U_c \cdot c}{4\pi \cdot R^2} \times K_{cs} \cdot \lambda_e \cdot U_e \cdot c\right]$$

and rearranging and simplifying

$$= \frac{[U_c \cdot c] \times [h \cdot K_{cs} \cdot U_e]}{4\pi \cdot R^2}$$

and substituting per equations *(27)* and *(21)* yields the result

(33) $$\text{Force} = \frac{Q_s \cdot Q_e}{4\pi \cdot R^2}$$

which is Coulomb's law as it naturally occurs.

If a constant of proportionality, `k`, is introduced to accommodate choice of the units of charge, and the constant $4\pi$ is absorbed into that new constant, then the result (using `q` for charge since the added constant requires an accordingly different variable) is

*(34)*
$$\text{Force} = k \cdot \frac{q_s \cdot q_e}{R^2}$$

which is Coulomb's Law as originally formulated.

Now, here, Coulomb's Law is not a law obtained by inference from empirical data as is the Coulomb's Law of traditional 20th Century physics. Rather, it is a derived result for which the derivation steps were:

    (a) The concept of field / wave and charge / center-of-oscillation.

    (b) Newton's `Acceleration = Force / Mass` is actually `Acceleration = [Wave (from the center-of-oscillation)] × [Responsiveness (of the center)]`.

    (c) Development of Responsiveness in terms of center and wave interaction.

    (d) Development of Wave in terms of center behavior and propagation characteristics.

    (e) Development of the relationship between the encountered center charge and its oscillation, and of the source center charge and its oscillation.

    (f) Combination of the above resulting in Coulomb's Law.

The success of the derivation and of the essentially quite simple analysis demonstrates that the effect we call inertial mass is merely a particle's effectiveness in intercepting a portion of incoming medium waves propagating from a charge.

## 5. *MASS AND THE LORENTZ CONTRACTIONS*

Now that it is clear that a charged particle [electron, proton, etc.] is a center-of-oscillation, oscillating per equation *(23)*; propagating an oscillatory wave of medium outward in all directions, also per equation *(23)*; and having a cross section for intercepting such waves from other centers per equation *(13)*; the effect of its velocity on the particle can be analyzed in terms of the behavior of those characteristics: the center oscillation, the propagated wave, and the cross section. The analysis must begin with Einstein's derivation of the mass-energy equivalence $[E = m \cdot c^2]$ and an important aspect that Einstein failed to note.

Kinetic energy, `KE`, is defined as the work done by the force, `f`, acting on the particle or object of mass, `m`, over the distance that the force acts, `s`. This quantity is calculated by integrating the action over differential distances.

*(35)*
$$KE = \int_0^s f \cdot ds \qquad \text{[Per above definition]}$$

$$= \int_0^s \frac{d(m \cdot v)}{dt} \cdot ds \qquad \text{[Newton's 2nd law]}$$

$$= \int_0^{(m \cdot v)} \frac{ds}{dt} \cdot d(m \cdot v) \qquad \text{[Rearrangement of form]}$$

$$= \int_0^{(m \cdot v)} v \cdot d(m \cdot v) \qquad [v = {}^{ds}/_{dt}]$$

$$= \int_0^v v \cdot d\left[\frac{m_r \cdot v}{\left[1 - \frac{v^2}{c^2}\right]^{\frac{1}{2}}}\right] \qquad [\text{m is } m_r \text{ Lorentz contracted by v}]$$

$$= \frac{m_r \cdot v^2}{\left[1 - \frac{v^2}{c^2}\right]^{\frac{1}{2}}} - m_r \cdot \int_0^v \left[\frac{v \cdot dv}{\left[1 - \frac{v^2}{c^2}\right]^{\frac{1}{2}}}\right] \qquad [\text{integration by parts}]$$

$$= \frac{m_r \cdot v^2}{\left[1 - \frac{v^2}{c^2}\right]^{\frac{1}{2}}} + m_r \cdot c^2 \cdot \left[1 - \frac{v^2}{c^2}\right]^{\frac{1}{2}} - m_r \cdot c^2 \qquad [\text{integration of 2}^{nd}\text{ term}]$$

$$KE = \frac{m_r \cdot c^2}{\left[1 - \frac{v^2}{c^2}\right]^{\frac{1}{2}}} - m_r \cdot c^2 \qquad [\text{rearrangement \& Simplification}]$$

$$= m_v \cdot c^2 - m_r \cdot c^2 \qquad [\text{Lorentz transform}]$$

$$= \{\text{Total Energy}\} - \{\text{Rest Energy}\}$$

$$\text{and} \quad E = m \cdot c^2 \quad \text{as a general principle}$$

Focused on the concept of kinetic energy as the excess of total energy over fixed rest energy, Einstein missed the more important concept of *Energy in Kinetic Form* and *Energy in Rest Form*. Proceeding slightly differently in the above derivation starting from the step at "*[integration of 2$^{nd}$ term]*" the following are obtained.

(36)
$$KE = \frac{m_r \cdot v^2}{\left[1 - \frac{v^2}{c^2}\right]^{\frac{1}{2}}} + m_r \cdot c^2 \cdot \left[1 - \frac{v^2}{c^2}\right]^{\frac{1}{2}} - m_r \cdot c^2 \qquad [\text{"integration of 2}^{nd}\text{ term" result here repeated}]$$

$$KE + m_r \cdot c^2 = \frac{m_r}{\left[1 - \frac{v^2}{c^2}\right]^{\frac{1}{2}}} \cdot v^2 + m_r \cdot \left[1 - \frac{v^2}{c^2}\right]^{\frac{1}{2}} \cdot c^2 \qquad [\text{rearranging}]$$

$$\begin{bmatrix}\text{Total} \\ \text{Energy}\end{bmatrix} = \begin{bmatrix}\text{Energy in} \\ \text{Kinetic Form}\end{bmatrix} + \begin{bmatrix}\text{Energy in} \\ \text{Rest Form}\end{bmatrix}$$

Considering and evaluating the three terms of equation *(36)*:

[a] Left Term:  $KE + m_r \cdot c^2$ = Kinetic plus Rest energies = Total Energy
   = Total Mass when divided by $c^2$.

[b] Right Term:

$$m_r \cdot \left[1 - \frac{v^2}{c^2}\right]^{1/2} \cdot c^2$$

Energy in Rest Form
= A Lorentz contracted reduced rest energy -- corresponding to the part of the center's oscillation that is in "rest form", i.e. spherically symmetrical.

= $m_r$ when $v = 0$.

= "Mass in Rest Form" when divided by $c^2$.

[c] Middle Term:

$$\frac{m_r}{\left[1 - \frac{v^2}{c^2}\right]^{1/2}} \cdot v^2$$

Energy in Kinetic Form
= An enhanced kinetic energy -- corresponding to the part of the center's oscillation that is distorted in the direction of motion, i.e. symmetrical cylindrically relative to the motion vector.

= $0$ when $v = 0$.

= "Mass in Kinetic Form" when divided by $c^2$.

therefore:

*(37)*  $\begin{bmatrix}\text{Total}\\ \text{Mass}\end{bmatrix} = \begin{bmatrix}\text{Mass in}\\ \text{Kinetic Form}\end{bmatrix} + \begin{bmatrix}\text{Mass in}\\ \text{Rest Form}\end{bmatrix}$

and it is those quantities that are to be treated in terms of the center-of-oscillation's responsiveness as a function of its oscillation and the corresponding wave.

To describe the behavior of the center and the various differences in the propagated waves in different directions from the center, the propagation is modeled resolved into three components: forward, rearward, and sideward. The directions are all relative to the direction of the center's velocity as depicted in Figure 4, below. For purposes of analysis these orthogonal components represent the propagated wave in all directions. The wave in any particular direction is the resultant of that directions' projection on the forward or rearward component (whichever is at a nearer angle) and on the sideward component.

For a center at absolute rest propagation of waves is the same in all directions at velocity $c = \lambda_r \cdot f_r$. See Figure 4, below. (In the figure the "up", "down", "left" and "right" are all "sideward".)

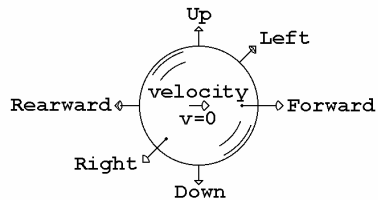

*Figure 4*
*A Center of Oscillation at Rest*

The effects that occur when this center is moving at velocity $v$ are described in two steps: the first to address the *Mass in Rest Form* [which, per equation *(36)*, <u>decreases</u> as $v$ increases] and the second to address the *Mass in Kinetic Form* [which, per equation *(36)*, <u>increases</u> as $v$ increases].

<u>*Step #1*</u>

The center's rest frequency decreases and its rest wavelength correspondingly increases, the product still being $c$.

(38)
$$f_v = f_r \cdot \left[1 - \frac{v^2}{c^2}\right]^{1/2} \quad \text{[Center frequency decreases]}$$

$$\lambda_v = \lambda_r \cdot \frac{1}{\left[1 - \frac{v^2}{c^2}\right]^{1/2}} \quad \text{[Center wavelength increases]}$$

$$v_{wave} = \lambda_v \cdot f_v = \lambda_r \cdot f_r = c \quad \text{[Wave velocity remains at } c\text{]}$$

$$m'_r = m_r \cdot \left[\frac{f_v}{f_r}\right] = m_r \cdot \left[1 - \frac{v^2}{c^2}\right]^{1/2} \quad \text{[Mass in Rest Form becomes less than } v = 0 \text{ rest mass]}$$

This has the effect of decreasing the amount of rest mass since *mass ∝ frequency ∝ 1/wavelength*. However Step #1 is the first of two steps of analysis of an actual event that occurs as one whole, therefore these results are interim, pending Step #2.

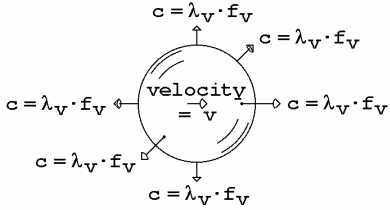

*Figure 5*
*A Center of Oscillation at Step #1*

<u>*Step #2*</u>

As presented in conjunction with equations *(1)* and *(2)*, when it is in motion the center of oscillation must adjust the velocity at which it propagates in the forward and rearward directions. To maintain propagated wave velocity at $c$ in the direction of center motion, $v$, the wave must be actually propagated forward by the center at $c'=c-v$ relative to the center itself so that the wave velocity relative to at absolute rest remains $c$. While the center can oscillate at only one frequency, it can propagate at different wavelengths in different directions. To propagate forward at $c'$ while maintaining the frequency at the $f_v$ of equation *(38)* requires that the wavelength change to a smaller value, $\lambda_{fwd}$.

Likewise, rearward the wave must be actually propagated by the center at $c''=c+v$ relative to the center and, therefore with a greater wavelength, $\lambda_{rwd}$.

As the center "sees" it, Figure 6 next page, the center is oscillating at $f_v$, with the wavelength (as always) set by the propagation velocity in any particular direction.

As at absolute rest "sees" it, Figure 7 next page, the center is propagating different forward and rearward frequencies, $f_{fwd}$ and $f_{rwd}$.

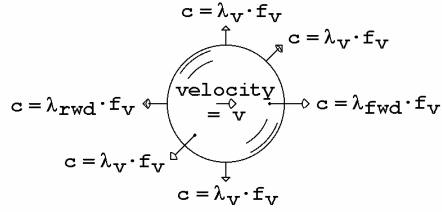

*Figure 6*
*A Center of Oscillation at Step #2*
*as the Center "Sees" It*

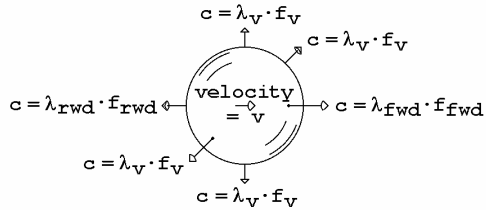

*Figure 7*
*A Center of Oscillation at Step #2*
*as at Absolute Rest "Sees" It*

In the preceding figures the "$rwd$" (rearward) and "$fwd$" (forward) wavelengths and frequencies are as follows,

*(39)*

$$\lambda_{fwd} = \frac{c'}{f_v} = \frac{c-v}{f_v} = \frac{c \cdot \left[1 - \frac{v}{c}\right]}{f_r \cdot \left[1 - \frac{v^2}{c^2}\right]^{\frac{1}{2}}} = \lambda_r \cdot \frac{\left[1 - \frac{v}{c}\right]^{\frac{1}{2}}}{\left[1 + \frac{v}{c}\right]^{\frac{1}{2}}}$$

$$= \lambda_r \cdot \left[\frac{c-v}{c+v}\right]^{\frac{1}{2}}$$

$$f_{fwd} = \frac{c}{\lambda_{fwd}} = f_r \cdot \left[\frac{c+v}{c-v}\right]^{\frac{1}{2}}$$

$$\lambda_{rwd} = \lambda_r \cdot \left[\frac{c+v}{c-v}\right]^{\frac{1}{2}}$$

$$f_{rwd} = f_r \cdot \left[\frac{c-v}{c+v}\right]^{\frac{1}{2}}$$

and the propagated waves travel at $c$ in all directions as observed by the center doing the propagating and as observed from at absolute rest.

So far the development has demonstrated a <u>decrease</u> in rest mass per equation *(38)*. Actually the total mass <u>increases</u> by the same factor as the Lorentz Transforms describe. To analyze how this occurs requires returning to the details of the interaction of the arriving wave and the center's responsiveness as was done above in deriving Coulomb's Law.

The analysis here is of those factors affecting the interaction relative to Step #1, above, having already occurred that is, the starting point is the center with reduced rest mass, $m'_r$, reduced oscillation frequency, $f_v$, and increased wavelength, $\lambda_v$ of equation *(38)*. The procedure is to investigate the change in the center's responsiveness due to its motion, the change as viewed from the direction of each of the orthogonal components of the oscillation.

Referring back to Newton's Law as restated, equation *(7)* repeated below, the responsiveness part of that equation is equation *(12)*, also repeated below.

*(7)*
$$\begin{bmatrix} \text{Acceleration} \\ \text{Resulting} \end{bmatrix} = \begin{bmatrix} \text{Wave} \\ \text{Potential} \\ \text{Impulse} \end{bmatrix} \times \begin{bmatrix} \text{Responsiveness} \\ \text{of the Center} \end{bmatrix}$$

Acceleration = Wave × Responsiveness.

*(12)* Responsiveness = [ Cross-section ]·[Amplitude]·[Frequency]

$$= [\, K_{cs} \cdot \lambda_e^2 \,] \cdot [\, U_e \,] \cdot [\, f_e \,]$$

where: $K_{cs}$ = a constant for the proportionality
$\lambda_e$ = the wavelength of the encountered center oscillation
$U_e$ = its amplitude, and
$f_e$ = its frequency

Now, at center velocity $v$, in the forward direction the wave is propagated by the moving center at $c' = c - v$ so that the propagation velocity, $c$, is changed by the factor $[c-v]/c$, a reduction by the factor of equation *(40)* below.

*(40)*
$$\text{factor} = \frac{c'}{c} = \frac{c-v}{c} = 1 - \frac{v}{c}$$

Since the propagated wave is reduced by that factor in the forward direction, the effective amplitude, $U_s$, of its "source" center in that direction must also be so reduced. And for that center in its "encountered" center role that constitutes a reduction by the same factor of equation *(39)* in $U_e$, the second of the three factors, cross-section, amplitude and repetition rate, in the center's responsiveness, equation *(12)*.

In addition, because the encountered center is now moving at velocity v, in the forward direction, toward incoming waves from a source center, the repetition rate, the third of the three factors in the equation *(12)* expression for responsiveness of wave center interaction, is increased. That is, it is increased by the factor $[c+v]/c$, the factor of equation *(41)*, below.

*(41)*
$$\frac{c+v}{c} = 1 + \frac{v}{c}$$

The combination of those two factors changes the at rest responsiveness by a factor equal to their product, that is by

*(42)*
$$\text{product of factors} = \left[1 - \frac{v}{c}\right] \cdot \left[1 + \frac{v}{c}\right]$$

$$= \text{change in responsiveness}$$

$$= 1 - \frac{v^2}{c^2}$$

a reduction in responsiveness, an increase in mass. Exactly analogous reasoning for the rearward direction results in the same overall change factor as equation *(42)*.

For the direction of each of the four components to the side (up, down, right, and left in the above figures) the case of interaction with waves coming in at right angles to the direction of motion of the center, the repetition rate is unchanged. There is no motion of the center toward or away from the incoming waves so there is no factor for a change in repetition rate due to such an effect. (The repetition rate is less due to the reduction of the center's frequency from $f_r$ to $f_v$ at Step #1, equation *(38)*, but that has already been accounted for in Step #1 and the current changes are relative to the results of Step #1.)

For this sideward case the cross-section is no longer a circle, however. In the forward direction the at rest circle's radius has become $\lambda_{fwd}$ instead of $\lambda_v$ and in the rearward direction $\lambda_{rwd}$ instead of $\lambda_v$.

The change factors are, from equation *(38)*:

*(43)*
$$\lambda_{fwd} = \frac{c-v}{f_v} = \frac{c \cdot \left[1 - \frac{v}{c}\right]}{f_v} = \lambda_v \cdot \left[1 - \frac{v}{c}\right]$$

$$\frac{\lambda_{fwd}}{\lambda_v} = 1 - \frac{v}{c}$$

*(44)*
$$\frac{\lambda_{rwd}}{\lambda_v} = 1 + \frac{v}{c}$$

so that the product of the change factors is, once again, equation *(42)*.

Therefore, from every direction the Step #2 responsiveness is reduced by the factor of equation *(42)* due to the motion of the center at velocity $v$. The mass is accordingly increased by the reciprocal of that factor. These changes are relative to the conditions at the end of Step #1, where the mass of the center, there $m'_r$, was as in equation *(38)*. Therefore $m_v$, the center overall mass after Step #2, its mass at velocity $v$, is

*(45)*
$$m_v = \begin{bmatrix} \text{Mass } m'_r \text{ of} \\ \text{equation} \\ 38 \end{bmatrix} \times \begin{bmatrix} \frac{1}{\text{Factor of equation 41}} \end{bmatrix}$$

$$= \left[ m_r \cdot \left[1 - \frac{v^2}{c^2}\right]^{\frac{1}{2}} \right] \times \left[ \frac{1}{1 - \frac{v^2}{c^2}} \right]$$

$$= m_r \cdot \frac{1}{\left[1 - \frac{v^2}{c^2}\right]^{\frac{1}{2}}}$$

which is the same as that called for by the Lorentz Contraction.

### *CONCLUSION*

The physical effect that we refer to as inertial mass, its behavior in Newton's Laws of Motion and its relativistic behavior according to the Lorentz Contractions are all properly interpreted in terms of particles' behavior as propagators of waves and as interceptors of such waves propagated by other such particles. Inertial mass is the extent of the ability of the particle's cross section, amplitude and frequency to intercept and interact with such waves.